\documentclass[copyright,noncommercial]{eptcs}

\providecommand{\event}{}
\providecommand{\volume}{}
\providecommand{\anno}{2009}
\providecommand{\firstpage}{1}
\usepackage{amsmath,latexsym,amsfonts,amssymb,amsthm}
\usepackage{rotating,graphs,stmaryrd}

\theoremstyle{definition}

\theoremstyle{remark}
\newtheorem{example}{Example}

\renewcommand{\L}{\mathcal{L}}
\newcommand{\Z}{\mathbb{Z}}
\newcommand{\B}{\mathbb{B}}
\newcommand{\C}{\mathrm{Char}}
\newcommand{\I}{\mathrm{I}}

\newcommand{\Sem}[1]{\llbracket{#1}\rrbracket}

\newcommand{\G}{\mathcal{G}}
\newcommand{\R}{\mathcal{R}}

\newcommand{\var}{\mathrm{Var}}
\newcommand{\type}{\mathrm{type}}
\newcommand{\dom}{\mathrm{Dom}}
\newcommand{\failrm}{\mathrm{fail}}
\newcommand{\failtt}{\mathtt{fail}}
\newcommand{\skiptt}{\mathtt{skip}}

\newcommand{\tuple}[1]{\langle#1\rangle}

\newcommand{\dder}{\Rightarrow}

\newcommand{\DSto}{\mathop{\to}\limits}

\newcommand{\ifte}[3]{\mathtt{if}\ #1\ \mathtt{then}\ #2\ \mathtt{else}\ #3}
\newcommand{\ift}[2]{\mathtt{if}\ #1\ \mathtt{then}\ #2}

\newcommand{\fail}{\mathrm{fail}}

\title{The Semantics of Graph Programs}

\author{Detlef Plump 
\institute{Department of Computer Science\\
           The University of York, UK}
\and Sandra Steinert
\institute{Department of Computer Science\\
           The University of York, UK}
}

\def\titlerunning{Semantics of Graph Programs}
\def\authorrunning{D. Plump \& S. Steinert}

\begin{document}
\maketitle\thispagestyle{empty}

\begin{abstract}
GP (for Graph Programs) is a rule-based, nondeterministic programming language for solving graph problems at a high level of abstraction, freeing programmers from handling low-level data structures. The core of GP consists of four constructs: single-step application of a set of conditional graph-transformation rules, sequential composition, branching and iteration. We present a formal semantics for GP in the style of structural operational semantics. A special feature of our semantics is the use of \emph{finitely failing}\/ programs to define GP's powerful branching and iteration commands. 
\end{abstract}

\section{Introduction}
\label{sec:introduction}

This paper defines the semantics of GP, an experimental nondeterministic programming language for high-level problem solving in the domain of graphs. The language is based on conditional rule schemata for graph transformation (introduced in \cite{Plump-Steinert04a}) and thereby frees programmers from handling low-level data structures for graphs. The prototype implementation of GP compiles graph programs into bytecode for the York abstract machine, and comes with a graphical editor for programs and graphs \cite{Manning-Plump08b}.

GP has a simple syntax as its core contains only four commands: single-step application of a set of rule schemata, sequential composition, branching and as-long-as-possible iteration. Despite its simplicity, GP is computationally complete in that every computable function on graphs can be programmed \cite{Habel-Plump01a}. A major goal of the GP project is the development of a practical graph-transformation language that comes with a concise formal semantics, to facilitate program verification and other formal reasoning on programs. Also, a formal semantics provides implementors with a rigorous definition of the language that does not depend on a compiler or machine.


To define the meaning of GP programs, we adopt Plotkin's method of structural operational semantics \cite{Plotkin04a}. This approach is well established for imperative programming languages \cite{Nielson-Nielson07a} but is novel in the field of graph transformation. In brief, the method consists in devising inference rules which inductively define the effect of commands on program states. Whereas a classic state consists of the values of all program variables at a certain point in time, the analogue for graph transformation is the graph on which the rules of a program operate.

As GP is nondeterministic, our semantics assigns to a program $P$ and an input graph $G$ \emph{all}\/ graphs that can result from executing $P$ on $G$. A special feature of the semantics is the use of failing computations to define powerful branching and iteration constructs. (Failure occurs when a set of rule schemata to be executed is not applicable to the current graph.) While the conditions of branching commands in traditional programming languages are boolean expressions, GP uses arbitrary programs as conditions. The evaluation of a condition $C$ succeeds if there \emph{exists} an execution of $C$ on the current graph that produces a graph. On the other hand, the evaluation of $C$ is unsuccessful if all executions of $C$ on the current graph result in failure. In this case $C$ \emph{finitely fails} on the current graph.

In logic programming, finite failure (of SLD resolution) is used to define negation \cite{Clark78a}. In the case of GP, it allows to ``hide'' destructive executions of the condition $C$ of a statement $\ifte{C}{P}{Q}$. This is because after evaluating $C$, the resulting graph is discarded and either $P$\/ or $Q$\/ is executed on the graph with which the branching statement was entered. Finite failure also allows to elegantly lift the application of as-long-as-possible iteration from sets of rule schemata (as in \cite{Plump-Steinert04a}) to arbitrary programs: the body of a loop can no longer be applied if it finitely fails on the current graph.

Control constructs which allow programmers to write ``strategies'' for applying rewrite rules have long been present in term-rewriting languages such as Elan \cite{Borovansky-Kirchner-Kirchner-Moreau02a} and Stratego \cite{Bravenboer-van_Dam-Olmos-Visser06a}. These languages allow recursive definitions of strategies whereas GP is based on a small set of built-in, non-recursive constructs. (See \cite{Steinert07a} for an extension of GP with recursive procedures.) 

Another difference between GP and languages such as Elan and Stratego is that strategies in the latter languages rely on the structure of the objects that they manipulate, that is, on the tree structure of terms. In both languages, term-rewrite rules are applied at the root of a term so that traversal operations are needed to apply rules and strategies deep inside terms. In contrast, the semantics of GP's control constructs does not depend on the structure of graphs and is completely orthogonal to the semantics of rule schemata. This provides a clear separation of concerns between rules and the control of rules, making it easy to adapt GP's semantics to different formats of rules or graphs.\footnote{In the extreme, one could even replace the underlying formalism of graph-transformation with some other rule-based framework, such as string or term rewriting.}

The contributions of this paper can be summarised as follows:
\begin{itemize}
\item A graph-transformation language with \emph{simple} syntax and semantics, facilitating understanding by programmers and formal reasoning on programs. Our experience so far is that very often short and easy to understand programs can be written to solve problems on graphs (see \cite{Plump09a} for various small case studies).
\item The first formal operational semantics for a graph-transformation language (to the best of our knowledge). Well-known languages such as AGG \cite{Ermel-Rudolf-Taentzer99a}, Fujaba \cite{Nickel-Niere-Zuendorf00a} and GrGen \cite{Geiss-Batz-Grund-Hack-Szalkowski06a} have no formal semantics. The only graph-transformation language with a complete formal semantics that we are aware of is PROGRES \cite{Schuerr-Winter-Zuendorf99a}. Its semantics, given by Sch{\"u}rr in his dissertation \cite{Schuerr91a}, translates programs into control-flow diagrams and consists of more than 300 rules (including the definition of the static semantics) .
\item A powerful branching construct based on the concept of finite failure, allowing to conveniently express complex destructive tests on input graphs. In addition, finite failure enables an elegant definition of as-long-as-possible iteration. These definitions do not depend on the structure of graphs and can be used for string- or term-based rewriting languages, too. 
\end{itemize}

The rest of this paper is structured as follows. The next section reviews the graph-trans\-for\-ma\-tion formalism underlying GP, the so-called double-pushout approach with relabelling. Section \ref{sec:rule_schemata} introduces conditional rule schemata as the building blocks of GP programs. In Section \ref{sec:programs}, we discuss an example program for graph colouring and define the abstract syntax of graph programs. Section \ref{sec:semantics} presents our formal semantics of GP in the style of structural operational semantics. In Section \ref{sec:conclusion}, we conclude and mention some topics for future work.

\section{Graph Transformation}
\label{sec:gratra}

\enlargethispage{\baselineskip}

We briefly review the model of graph transformation underlying GP, the double-pushout approach with relabelling \cite{Habel-Plump02c}. Our presentation is tailored to GP in that we consider graphs over a fixed label alphabet, and rules in which only the interface may contain unlabelled nodes.

GP programs operate on graphs labelled with sequences of integers and strings. (The reason for using sequences will become clear in Section \ref{sec:programs}.) To formalise this, let $\Z$ be the set of integers and $\C$ be a finite set of characters---we may think of $\C$ as the characters that can be typed on a keyboard. We fix the label alphabet $\L = (\Z \cup \C^*)^+$ consisting of all nonempty sequences made up from integers and character strings. 

A \emph{partially labelled graph} over $\L$ (or \emph{graph}\/ for short) is a system $G=(V_G,E_G,s_G,t_G,l_G,m_G)$, where $V_G$ and $E_G$ are finite sets of \emph{nodes} (or \emph{vertices}) and \emph{edges}, $s_G,t_G\colon E_G\rightarrow V_G$ are the \emph{source} and \emph{target} functions for edges, $l_G\colon V_G\to \L$ is the partial node labelling function and $m_G\colon E_G\to \L$ is the (total) edge labelling function. Given a node $v$, we write $l_G(v)=\perp$ to express that $l_G(v)$ is undefined. Graph $G$ is \emph{totally labelled} if $l_G$ is a total function.  

The set of all totally labelled graphs over $\L$ is denoted by $\G$. GP programs operate on the graphs in $\G$, unlabelled nodes occur only in the interfaces of rules (see below) and are necessary in the double-pushout approach to relabel nodes. There is no need to relabel edges as they can always be deleted and reinserted with changed labels.

A \emph{graph morphism} $g\colon G\rightarrow H$ between graphs $G$ and $H$\/ consists of two functions $g_V\colon V_G\rightarrow V_H$ and $g_E\colon E_G\rightarrow E_H$\/ that preserve sources, targets and labels (that is, $s_H\circ g_E=g_V\circ s_G$, $t_H\circ g_E=g_V\circ t_G$, $m_H \circ g_E = m_G$, and $l_H(g(v))=l_G(v)$ for all $v$ such that $l_G(v) \neq \perp$). Morphism $g$ is an \emph{inclusion} if $g(x)=x$ for all nodes and edges $x$. It is \emph{injective}\/ if $g_V$ and $g_E$ are injective. 


A \emph{rule} $r = (L \gets K \to R)$ consists of two inclusions $K \to L$\/ and $K \to R$\/ where $L$\/ and $R$\/ are totally labelled graphs. Graph $K$\/ is the \emph{interface} of $r$. Intuitively, an application of $r$ to a graph will remove the items in $L - K$, preserve $K$, add the items in $R - K$, and relabel the unlabelled nodes in $K$. 
Given a graph $G$\/ in $\G$, an injective graph morphism $g\colon L \to G$\/ is a \emph{match} for $r$ if it satisfies the \emph{dangling condition}: no node in $g(L)-g(K)$ is incident to an edge in $G-g(L)$. In this case $G$ \emph{directly derives} the graph $H$\/ in $\G$ that is constructed from $G$\/ as follows:\footnote{See \cite{Habel-Plump02c} for an equivalent definition by graph pushouts.} 
\enlargethispage{\baselineskip}
\begin{enumerate}
\item Remove all nodes and edges in $g(L)-g(K)$.
\item Add disjointly all nodes and edges from $R-K$, keeping their labels. For $e \in E_R-E_K$, $s_H(e)$ is $s_R(e)$ if $s_R(e) \in V_R-V_K$, otherwise $g_V(s_R(e))$. Targets are defined analogously.  
\item For each node $v$ in $K$ with $l_K(v) = \bot$, $l_H(g_V(v))$ becomes $l_R(v)$. 
\end{enumerate}
We write $G \dder_{r,g} H$ (or just $G \dder_r H$) if $G$\/ directly derives $H$\/ as above.

Figure \ref{fig:direct_derivation} shows an example of a direct derivation. The rule in the upper row is applied to the left graph of the lower row, resulting in the right graph of the lower row. For simplicity, we do not depict edge labels and assume that they are all the same. The node identifiers 1 and 2 in the rule specify the inclusions of the interface. The middle graph of the lower row is an intermediate result (omitted in the above construction). This diagram represents a double-pushout in the category of partially labelled graphs over $\L$.

\begin{figure}[htb]
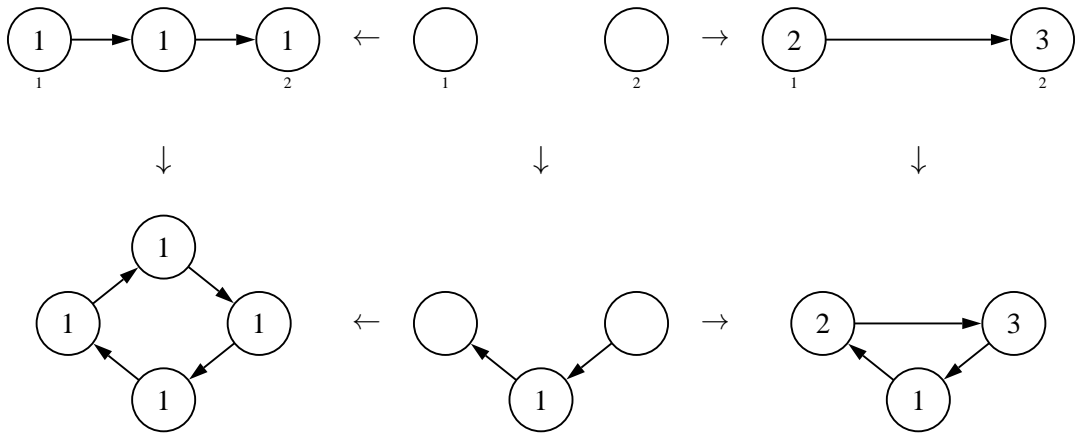

 \begin{center}
     \setlength{\unitlength}{0.5in}
   \graphnodesize{0.68}
   \grapharrowlength{0.2}
   \grapharrowwidth{0.6}
   \autodistance{1.0}
   \graphnodecolour{1}

\begin{tabular}{c}
\begin{graph}(3.4,1.5)(0,.95)
 \roundnode{n1}(.4,1.7)\autonodetext{n1}{1}\autonodetext{n1}[s]{\tiny 1}
 \roundnode{n2}(1.7,1.7)\autonodetext{n2}{1}
 \roundnode{n3}(3,1.7)\autonodetext{n3}{1}\autonodetext{n3}[s]{\tiny 2}
 \diredge{n1}{n2}
 \diredge{n2}{n3}
\end{graph}
\begin{graph}(.7,1.5)
 \freetext(.35,.75){$\gets$}
\end{graph}
\begin{graph}(2.8,1.5)(0,.95)
 \roundnode{n1}(.4,1.7)\autonodetext{n1}[s]{\tiny 1}
 \roundnode{n3}(2.4,1.7)\autonodetext{n3}[s]{\tiny 2}
\end{graph}
\begin{graph}(.7,1.5)
 \freetext(.35,.75){$\to$}
\end{graph}
\begin{graph}(3.4,1.5)(0,.95)
 \roundnode{n1}(.4,1.7)\autonodetext{n1}{2}\autonodetext{n1}[s]{\tiny 1}
 \roundnode{n3}(3,1.7)\autonodetext{n3}{3}\autonodetext{n3}[s]{\tiny 2}
 \diredge{n1}{n3}
\end{graph}
\\
\begin{graph}(3.4,.8)
 \freetext(1.7,.4){\large $\downarrow$}
\end{graph}
\begin{graph}(.7,.8)
\end{graph}
\begin{graph}(2.8,.8)
 \freetext(1.4,.4){\large $\downarrow$}
\end{graph}
\begin{graph}(.7,.8)
\end{graph}
\begin{graph}(3.4,.8)
 \freetext(1.7,.4){\large $\downarrow$}
\end{graph}
\\
\begin{graph}(3.4,2.4)(.4,0)
 \roundnode{n1}(1.1,1.2)\autonodetext{n1}{1}
 \roundnode{n2}(2.1,2)\autonodetext{n2}{1}
 \roundnode{n3}(3.1,1.2)\autonodetext{n3}{1}
 \roundnode{n4}(2.1,.4)\autonodetext{n4}{1}
 \diredge{n1}{n2}
 \diredge{n2}{n3}
 \diredge{n3}{n4}
 \diredge{n4}{n1}
\end{graph}
\begin{graph}(.7,2.4)
 \freetext(.35,1.2){$\gets$}
\end{graph}
\begin{graph}(2.8,2.4)(.7,0)
 \roundnode{n1}(1.1,1.2)
 \roundnode{n3}(3.1,1.2)
 \roundnode{n4}(2.1,.4)\autonodetext{n4}{1}
 \diredge{n3}{n4}
 \diredge{n4}{n1}
\end{graph}
\begin{graph}(.7,2.4)
 \freetext(.35,1.2){$\to$}
\end{graph}
\begin{graph}(3.4,2.4)(.4,0)
 \roundnode{n1}(1.1,1.2)\autonodetext{n1}{2}
 \roundnode{n3}(3.1,1.2)\autonodetext{n3}{3}
 \roundnode{n4}(2.1,.4)\autonodetext{n4}{1}
 \diredge{n1}{n3}
 \diredge{n3}{n4}
 \diredge{n4}{n1}
\end{graph}
\end{tabular}
 \end{center}
\caption{A direct derivation \label{fig:direct_derivation}}
\end{figure}

To define conditional rules, we equip rules with predicates that restrict sets of matches. A \emph{conditional rule} $q = (r,P)$ consists of a rule $r$ and a predicate $P$ on graph morphisms. Given totally labelled graphs $G$, $H$\/ and a match $g\colon L \to G$ for $q$, we write $G \dder_{q,g} H$\/ (or just $G \dder_q H$) if $P(g)$ holds and $G \dder_{r,g} H$. For a set of conditional rules $\R$, we write $G \dder_{\R} H$ if there is some $q$ in $\R$ such that $G \dder_q H$.


\section{Conditional Rule Schemata}
\label{sec:rule_schemata}

A GP program is essentially a list of declarations of conditional rule schemata together with a command sequence for controlling the application of the schemata. Rule schemata generalise rules in that labels can contain expressions over parameters of type integer or string. In this section, we give an abstract syntax for the textual components of conditional rule schemata and interpret them as sets of conditional rules. 

Figure \ref{fig:condruleschema} shows an example for the declaration of a conditional rule schema. It consists of the identifier \texttt{bridge} followed by the declaration of formal parameters, the left and right graphs of the schema which are labelled with expressions over the parameters, the node identifiers $\mathtt{1}$, $\mathtt{2}$, $\mathtt{3}$ determining the interface of the schema, and the keyword \texttt{where} followed by the condition.

\begin{figure}[htb]
 \begin{center}
  \input{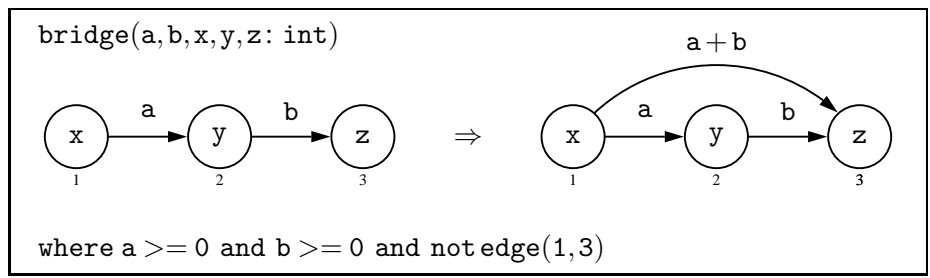}
 \end{center}
\caption{A conditional rule schema}\label{fig:condruleschema}
\end{figure}

In the GP programming system \cite{Manning-Plump08b}, rule schemata are constructed with a graphical editor. Figure \ref{fig:label_syntax} gives a grammar in Extended Backus-Naur Form for node and edge labels in the left and right graph of a rule schema (categories LeftLabel and RightLabel).\footnote{The grammars in Figure \ref{fig:label_syntax} and Figure \ref{fig:condition_syntax} are ambiguous, we use parentheses to disambiguate expressions where necessary.} 
Labels can be sequences of expressions separated by underscores, as will be demonstrated by Example \ref{ex:2-colouring} in Section \ref{sec:programs}. We require that labels in the left graph must be simple expressions because their values at execution time are determined by graph matching. All variable identifiers in the right graph must also occur in the left graph. Every expression in category $\mathrm{Exp}$ has type $\mathtt{int}$ or $\mathtt{string}$, where arithmetical operators expect arguments of type $\mathtt{int}$ and the type of variable identifiers is determined by their declarations.
\begin{figure}[htb]
\renewcommand{\arraystretch}{1.2}
\begin{center}
\begin{tabular}{lcl}
LeftLabel & ::= & SimpleExp ['\_' LeftLabel] \\
RightLabel & ::= & Exp ['\_' RightLabel] \\
SimpleExp & ::= &['-'] Num $\mid$ String $\mid$ VarId \\
Exp & ::= & SimpleExp $\mid$ Exp ArithOp Exp \\
ArithOp & ::= & '\verb#+#' $\mid$ '\verb#-#' $\mid$ '$\ast$' $\mid$ '\verb#/#' \\
Num & ::= & Digit \{Digit\} \\
String & ::= & '\,''\,' \{Char\} '\,''\,'
\end{tabular}
\end{center}
\caption{Syntax of node and edge labels}\label{fig:label_syntax}
\end{figure}
\begin{figure}[htb]
\renewcommand{\arraystretch}{1.2}
\begin{center}
\begin{tabular}{lcl}
BoolExp & ::= & \texttt{edge} '(' Node ',' Node ')' $\mid$ Exp RelOp Exp \\
&& $\mid$ \texttt{not} BoolExp $\mid$ BoolExp BoolOp BoolExp \\
Node & ::= & Digit \{Digit\} \\
RelOp & ::= & '\verb#=#' $\mid$ '\verb#\=#' $\mid$ '\verb#>#' $\mid$ '\verb#<#' $\mid$ '\verb#>=#' $\mid$ '\verb#<=#' \\
BoolOp & ::= & \texttt{and} $\mid$ \texttt{or} 
\end{tabular}
\end{center}
\caption{Syntax of conditions}\label{fig:condition_syntax}
\end{figure}

The condition of a rule schema is a boolean expression built from expressions of category Exp and the special predicate \texttt{edge}, see Figure \ref{fig:condition_syntax}. Again, all variable identifiers occurring in the condition must also occur in the left graph of the schema. The predicate \texttt{edge} demands the (non-)existence of an edge between two nodes in the graph to which the rule schema is applied. For example, the expression $\mathtt{not\, edge(1,3)}$ in the condition of Figure \ref{fig:condruleschema} forbids an edge from node 1 to node 3 when the left graph is matched. 

We interpret a conditional rule schema as the (possibly infinite) set of conditional rules that is obtained by instantiating variables with any values and evaluating expressions. To define this, consider a declaration $D$ of a conditional rule-schema. Let $L$\/ and $R$\/ be the left and right graphs of $D$, and $c$ the condition. We write $\var(D)$ for the set of variable identifiers occurring in $D$. Given $x$ in $\var(D)$, $\type(x)$ denotes the type associated with $x$. An \emph{assignment} is a mapping $\alpha\colon \var(D) \to (\Z \cup \C^*)$ such that for each $x$ in $\var(D)$, $\type(x) = \mathtt{int}$ implies $\alpha(x) \in \Z$, and $\type(x) = \mathtt{string}$ implies $\alpha(x) \in \C^*$. 

Given a label $l$\/ of category $\mathrm{RightLabel}$ occuring in $D$ and an assignment $\alpha$, the value $l^{\alpha} \in \L$ is inductively defined. If $l$\/ is a numeral or a sequence of characters, then $l^{\alpha}$ is the integer or character string represented by $l$ (which is independent of $\alpha$). If $l$\/ is a variable identifier, then $l^{\alpha} = \alpha(l)$. Otherwise, $l^{\alpha}$ is obtained from the values of $l$'s components. If $l$\/ has the form $e_1 \oplus e_2$ with $\oplus$ in $\mathrm{ArithOp}$ and $e_1,e_2$ in $\mathrm{Exp}$, then $l^{\alpha} = e_1^{\alpha} \oplus_{\Z} e_2^{\alpha}$ where $\oplus_{\Z}$ is the integer operation represented by $\oplus$.\footnote{For simplicity, we consider division by zero as an implementation-level issue.} If $l$\/ has the form $e\_m$ with $e$ in $\mathrm{Exp}$ and $m$ in $\mathrm{RightLabel}$, then $l^{\alpha} = e^{\alpha}m^{\alpha}$ (the concatenation of $e^{\alpha}$ and $m^{\alpha}$). Note that our definition of $l^{\alpha}$ covers all labels in $D$ since $\mathrm{LeftLabel}$ is a subcategory of $\mathrm{RightLabel}$. 

The value of the condition $c$ in $D$ not only depends on an assignment but also on a graph morphism. For, if $c$ contains the predicate \texttt{edge}, we need to consider the structure of the graph to which we want to apply the rule schema. Consider an assignment $\alpha$ and let $L^{\alpha}$ be obtained from $L$ by replacing each label $l$ with $l^{\alpha}$. Let $g\colon L^{\alpha} \to G$ be a graph morphism with $G \in \G$. Then for each Boolean subexpression $b$ of $c$, the value $b^{\alpha,g}$ in $\B = \{\mathtt{tt,ff}\}$ is inductively defined. If $b$\/ has the form $e_1 \bowtie e_2$ with $\bowtie$ in $\mathrm{RelOp}$ and $e_1,e_2$ in $\mathrm{Exp}$, then $b^{\alpha,g} = \mathtt{tt}$ if and only if $e_1^{\alpha} \bowtie_{\Z} e_2^{\alpha}$ where $\bowtie_{\Z}$ is the relation on integers represented by $\bowtie$. If $b$\/ has the form $\mathop{\mathtt{not}} b_1$ with $b_1$ in $\mathrm{BoolExp}$, then $b^{\alpha,g} = \mathtt{tt}$ if and only if $b_1^{\alpha,g} = \mathtt{ff}$. If $b$\/ has the form $b_1 \oplus b_2$ with $\oplus$ in $\mathrm{BoolOp}$ and $b_1,b_2$ in $\mathrm{BoolExp}$, then $b^{\alpha,g} = b_1^{\alpha,g} \oplus_{\B} b_2^{\alpha,g}$ where $\oplus_{\B}$ is the Boolean operation on $\B$ represented by $\oplus$. A special case is given if $b$\/ has the form $\mathtt{edge}(v,w)$ where $v,w$ are identifiers of interface nodes in $D$. We then have
\[b^{\alpha,g} = \left\{\begin{array}{ll} \mathtt{tt} & \text{if there is an edge from $g(v)$ to $g(w)$,}\\ \mathtt{ff} & \text{otherwise.} \end{array}\right.\]

Let now $r$ be the rule-schema identifier associated with declaration $D$. 
For every assignment $\alpha$, let $r^{\alpha}=(L^{\alpha} \gets K \to R^{\alpha},\, P^{\alpha})$ be the conditional rule given as follows:
\begin{itemize}
\item $L^{\alpha}$ and $R^{\alpha}$ are obtained from $L$ and $R$ by replacing each label $l$ with $l^{\alpha}$.
\item $K$ is the discrete subgraph of $L$ and $R$ determined by the node identifiers for the interface, where all nodes are unlabelled.
\item $P^{\alpha}$ is defined by: $P^{\alpha}(g)$ if and only if $g$ is a graph morphism $L^{\alpha} \to G$ such that $G \in \G$ and $c^{\alpha,g} = \mathtt{tt}$.
\end{itemize}
The \emph{interpretation} of $r$ is the rule set $\I(r) = \{r^{\alpha} \mid \text{$\alpha$ is an assignment}\}$. For notational convenience, we sometimes denote the relation $\dder_{\I(r)}$ by $\dder_r$. Note that $\I(r)$ is a (possibly infinite) set of conditional rules in the sense of Section \ref{sec:gratra}, grounding rule schemata in the theory of the double-pushout approach with relabelling \cite{Habel-Plump02c}.

\enlargethispage{2\baselineskip}
For example, the upper rows of Figure \ref{fig:rule-schema_appl} show the rule schema $\mathtt{bridge}$ of Figure \ref{fig:condruleschema} (without condition) and its instance $\mathtt{bridge}^{\alpha}$, where $\alpha(\mathtt{x})=0$, $\alpha(\mathtt{y})=\alpha(\mathtt{z})=1$, $\alpha(\mathtt{a})=3$ and $\alpha(\mathtt{b})=2$. The condition $c$ of $\mathtt{bridge}$ evaluates to the predicate $P^{\alpha}$ which is true for a match $g$ of the left-hand graph if and only if there is no edge from $g(1)$ to $g(3)$. (The subexpressions $\mathtt{a>=0}$ and $\mathtt{b>=0}$ evaluate to \texttt{tt} and hence can be ignored.) The lower rows of Figure \ref{fig:rule-schema_appl} show an application of $\mathtt{bridge}^{\alpha}$ by a graph morphism satisfying $P^{\alpha}$.
\begin{figure}[htb]
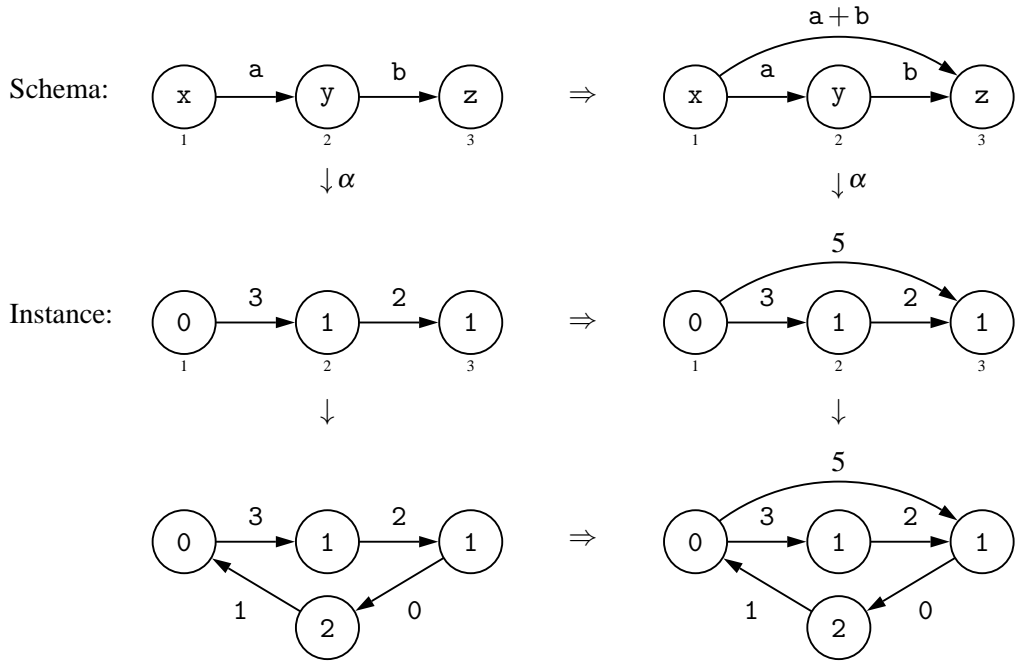

 \begin{center}
     \setlength{\unitlength}{0.5in}
   \graphnodesize{0.68}
   \grapharrowlength{0.2}
   \grapharrowwidth{0.6}
   \autodistance{1.0}
   \graphnodecolour{1}

\begin{tabular}{ll}
\raisebox{.56\unitlength}{Schema:} 
&
\begin{graph}(3.8,1.64)(.2,1.13)
 \roundnode{n1}(.6,1.7)
  \autonodetext{n1}{\texttt{x}}\autonodetext{n1}[s]{\tiny 1}
 \roundnode{n2}(2.1,1.7)
  \autonodetext{n2}{\texttt{y}}\autonodetext{n2}[s]{\tiny 2}
 \roundnode{n3}(3.6,1.7)
  \autonodetext{n3}{\texttt{z}}\autonodetext{n3}[s]{\tiny 3}
 \diredge{n1}{n2}\bowtext{n1}{n2}{.18}{\texttt{a}}
 \diredge{n2}{n3}\bowtext{n2}{n3}{.18}{\texttt{b}}
\end{graph}
\begin{graph}(1.4,1.64)
 \freetext(.7,.57){$\dder$}
\end{graph}
\begin{graph}(3.8,1.64)(.2,1.13)
 \roundnode{n1}(.6,1.7)
  \autonodetext{n1}{\texttt{x}}\autonodetext{n1}[s]{\tiny 1}
 \roundnode{n2}(2.1,1.7)
  \autonodetext{n2}{\texttt{y}}\autonodetext{n2}[s]{\tiny 2}
 \roundnode{n3}(3.6,1.7)
  \autonodetext{n3}{\texttt{z}}\autonodetext{n3}[s]{\tiny 3}
 \diredge{n1}{n2}\bowtext{n1}{n2}{.18}{\texttt{a}}
 \diredge{n2}{n3}\bowtext{n2}{n3}{.18}{\texttt{b}}
 \dirbow{n1}{n3}{.21}\bowtext{n1}{n3}{.28}{$\mathtt{a+b}$}
\end{graph}
\\
&
\begin{graph}(3.8,.5)
 \freetext(1.9,.3){$\downarrow$}\freetext(2.1,.3){\small$\alpha$}
\end{graph}
\begin{graph}(1.4,.5)
\end{graph}
\begin{graph}(3.8,.5)
 \freetext(1.9,.25){$\downarrow$}\freetext(2.1,.3){\small$\alpha$}
\end{graph}
\\
\raisebox{.56\unitlength}{Instance:} 
&
\begin{graph}(3.8,1.64)(.2,1.13)
 \roundnode{n1}(.6,1.7)
  \autonodetext{n1}{\texttt{0}}\autonodetext{n1}[s]{\tiny 1}
 \roundnode{n2}(2.1,1.7)
  \autonodetext{n2}{\texttt{1}}\autonodetext{n2}[s]{\tiny 2}
 \roundnode{n3}(3.6,1.7)
  \autonodetext{n3}{\texttt{1}}\autonodetext{n3}[s]{\tiny 3}
 \diredge{n1}{n2}\bowtext{n1}{n2}{.18}{\texttt{3}}
 \diredge{n2}{n3}\bowtext{n2}{n3}{.18}{\texttt{2}}
\end{graph}
\begin{graph}(1.4,1.64)
 \freetext(.7,.57){$\dder$}
\end{graph}
\begin{graph}(3.8,1.64)(.2,1.13)
 \roundnode{n1}(.6,1.7)
  \autonodetext{n1}{\texttt{0}}\autonodetext{n1}[s]{\tiny 1}
 \roundnode{n2}(2.1,1.7)
  \autonodetext{n2}{\texttt{1}}\autonodetext{n2}[s]{\tiny 2}
 \roundnode{n3}(3.6,1.7)
  \autonodetext{n3}{\texttt{1}}\autonodetext{n3}[s]{\tiny 3}
 \diredge{n1}{n2}\bowtext{n1}{n2}{.18}{\texttt{3}}
 \diredge{n2}{n3}\bowtext{n2}{n3}{.18}{\texttt{2}}
 \dirbow{n1}{n3}{.21}\bowtext{n1}{n3}{.28}{$5$}
\end{graph}
\\
&
\begin{graph}(3.8,.5)
 \freetext(1.9,.25){$\downarrow$}
\end{graph}
\begin{graph}(1.4,.5)
\end{graph}
\begin{graph}(3.8,.5)
 \freetext(1.9,.25){$\downarrow$}
\end{graph}
\\
&
\begin{graph}(3.8,2.3)(.2,.4)
 \roundnode{n1}(.6,1.7)
  \autonodetext{n1}{\texttt{0}}
 \roundnode{n2}(2.1,1.7)
  \autonodetext{n2}{\texttt{1}}
 \roundnode{n21}(2.1,.8)
  \autonodetext{n21}{\texttt{2}}
 \roundnode{n3}(3.6,1.7)
  \autonodetext{n3}{\texttt{1}}
 \diredge{n1}{n2}\bowtext{n1}{n2}{.18}{\texttt{3}}
 \diredge{n2}{n3}\bowtext{n2}{n3}{.18}{\texttt{2}}
 \diredge{n3}{n21}\bowtext{n3}{n21}{.18}{\texttt{0}}
 \diredge{n21}{n1}\bowtext{n21}{n1}{.18}{\texttt{1}}
\end{graph}
\begin{graph}(1.4,2.3)
 \freetext(.7,1.32){$\dder$}
\end{graph}
\begin{graph}(3.8,2.3)(.2,.4)
 \roundnode{n1}(.6,1.7)
  \autonodetext{n1}{\texttt{0}}
 \roundnode{n2}(2.1,1.7)
  \autonodetext{n2}{\texttt{1}}
 \roundnode{n21}(2.1,.8)
  \autonodetext{n21}{\texttt{2}}
 \roundnode{n3}(3.6,1.7)
  \autonodetext{n3}{\texttt{1}}
 \diredge{n1}{n2}\bowtext{n1}{n2}{.18}{\texttt{3}}
 \diredge{n2}{n3}\bowtext{n2}{n3}{.18}{\texttt{2}}
 \dirbow{n1}{n3}{.21}\bowtext{n1}{n3}{.28}{$5$}
 \diredge{n3}{n21}\bowtext{n3}{n21}{.18}{\texttt{0}}
 \diredge{n21}{n1}\bowtext{n21}{n1}{.18}{\texttt{1}}
\end{graph}
\end{tabular}
 \end{center}
\caption{Application of a rule schema using instantiation
         \label{fig:rule-schema_appl}}
\end{figure}

\section{Graph Programs}
\label{sec:programs}

We start by discussing an example program for graph colouring. 

\begin{example}[Computing a 2-colouring]
\label{ex:2-colouring}
A \emph{colouring} for a graph is an assignment of colours (integers) to nodes such that the source and target of each edge have different colours. A graph is \emph{2-colourable} (or \emph{bipartite}) if it possesses a colouring with at most two colours. The program \texttt{2-colouring} in Figure \ref{fig:2colouring} generates a 2-colouring for nonempty, connected input graphs without loops if such a colouring exists---otherwise the input graph is returned. The program consists of five rule-schema declarations, the \emph{macro} \texttt{colour} representing the rule-schema set $\mathtt{\{colour1,\, colour2\}}$, and the main command sequence following the key word \texttt{main}. 
\begin{figure}[htb]
 \begin{center}
  \input{Graphs/2colouring.prog}
 \end{center}
\caption{The program \texttt{2-colouring}}\label{fig:2colouring}
\end{figure}

Given an integer-labelled input graph, the program first uses the rule schema \texttt{choose} to pick any node and replace its label $x$ with $x\_0$. The underscore operator allows to add a \emph{tag} to a label, used here to add colours to labels. In general, a tagged label consists of a sequence of expressions joined by underscores.
After the first node has been coloured, the command \texttt{colour!} applies the rule schemata \texttt{colour1} and \texttt{colour2} nondeterministically as long as possible to colour all remaining nodes. In each iteration of the loop, an uncoloured node adjacent to an already coloured node $v$ gets the colour in $\{0,1\}$ that is complementary to $v$'s colour. If the input graph is connected, the graph resulting from \texttt{colour!} is correctly coloured if and only if the rule schema \texttt{illegal} is not applicable.  The latter is checked by the if-statement. If \texttt{illegal} is applicable, then the input must contain an undirected cycle of odd length and hence is not 2-colourable (see for example \cite{Kleinberg-Tardos06a}). In this case the loop \texttt{undo!} removes all tags to return the input graph unmodified. Note that the number of rule-schema applications performed by \texttt{2-colouring} is linear in the number of input nodes.

To make \texttt{2-colouring} applicable to graphs that are possibly empty or disconnected, we can insert a nested loop: 
\[ \text{\texttt{main = (choose; colour!)!; if illegal then undo!}.} \]
Now if the input graph is empty, \texttt{choose} fails which causes the outer loop to terminate and return the current (empty) graph. On the other hand, if the input consists of several connected components, the body of the outer loop is repeatedly called to colour each component. 
\end{example}

Figure \ref{fig:program_syntax} shows the abstract syntax of GP programs.\footnote{Where necessary we use parentheses to disambiguate programs.} A program consists of a number of declarations of conditional rule schemata and macros, and exactly one declaration of a main command sequence. 
The rule-schema identifiers (category RuleId) occurring in a call of category RuleSetCall refer to declarations of conditional rule schemata in category RuleDecl (see Section \ref{sec:rule_schemata}). Semantically, each rule-schema identifier $r$ stands for the set  $\I(r)$ of conditional rules induced by that identifier. A call of the form $\{r_1,\dots,r_n\}$ stands for the union $\bigcup_{i=1}^n\I(r_i)$.
\begin{figure}[htb]
\renewcommand{\arraystretch}{1.2}
\begin{center}
\begin{tabular}{lcl}
Prog & ::= & Decl \{Decl\} \\
Decl & ::= & RuleDecl $\mid$ MacroDecl $\mid$ MainDecl \\
MacroDecl & ::= & MacroId '=' ComSeq \\
MainDecl & ::= & \texttt{main} '=' ComSeq \\
ComSeq & ::= & Com \{';' Com\} \\
Com & ::= & RuleSetCall $\mid$ MacroCall \\
&& $\mid$ \texttt{if} ComSeq \texttt{then} ComSeq [\texttt{else} ComSeq] \\
&& $\mid$ ComSeq '!' \\
&& $\mid$ \texttt{skip} $\mid$ \texttt{fail} \\
RuleSetCall & ::= & RuleId $\mid$ '\{' [RuleId \{',' RuleId\}] '\}' \\
MacroCall & ::= & MacroId 
\end{tabular}
\end{center}
\caption{Abstract syntax of GP}\label{fig:program_syntax}
\end{figure}

Macros are a simple means to structure programs and thereby to make them more readable. Every program can be transformed into an equivalent macro-free program by replacing macro calls with their associated command sequences (recursive macros are not allowed). In the next section we use the terms ``program'' and ``command sequence'' synonymously, assuming that all macro calls have been replaced. 

The commands \texttt{skip} and \texttt{fail} can be expressed through the other commands (see next section), hence the core of GP includes only the call of a set of conditional rule schemata (RuleSetCall), sequential composition (';'), the if-then-else statement and as-long-as-possible iteration ('!'). 

\section{Semantics of Graph Programs}
\label{sec:semantics}

We present a formal semantics of GP in the style of Plotkin's structural operational semantics \cite{Plotkin04a}. As usual for this approach, inference rules inductively define a small-step transition relation $\to$ on \emph{configurations}. In our setting, a configuration is either a command sequence together with a graph, just a graph or the special element fail:
\[ \to \;\; \subseteq \; (\text{ComSeq} \times \G) \times 
      ((\text{ComSeq} \times \G) \cup \G \cup \{\fail\}). \]
Configurations in $\text{ComSeq} \times \G$ represent unfinished computations, given by a rest program and a state in the form of a graph, while graphs in $\G$ are proper results of computations. In addition, the element fail represents a failure state. A configuration $\gamma$ is \emph{terminal}\/ if there is no configuration $\delta$ such that $\gamma \to \delta$.

Each inference rule in Figure \ref{fig:core_sos_rules} consists of a premise and a conclusion separated by a horizontal bar. Both parts contain meta-variables for command sequences and graphs, where $R$ stands for a call in category RuleSetCall, $C,P,P',Q$ stand for command sequences in category ComSeq and $G,H$\/ stand for graphs in $\G$. Given a rule-set call $R$, let $\I(R) = \bigcup\{\I(r) \mid  \text{$r$ is a rule-schema identifier in $R$}\}$ (see Section \ref{sec:rule_schemata} for the definition of $\I(r)$). The \emph{domain} of $\dder_{\I(R)}$, denoted by $\dom(\dder_{\I(R)})$, is the set of all graphs $G$ in $\G$ such that $G \dder_{\I(R)} H$\/ for some graph $H$. Meta-variables are considered to be universally quantified. For example, the rule $\mathrm{[Call_1]}$ should be read as: ``For all $R$ in $\mathrm{RuleSetCall}$ and all $G,H$ in $\G$, $G \dder_{\I(R)} H$\/ implies $\tuple{R,\, G} \to H$.''

Figure \ref{fig:core_sos_rules} shows the inference rules for the core constructs of GP. We write $\to^+$ and $\to^*$ for the transitive and reflexive-transitive closures of $\to$. A command sequence $C$ \emph{finitely fails} on a graph $G \in \G$ if (1) there does not exist an infinite sequence $\tuple{C,\, G} \to \tuple{C_1,\, G_1} \to \dots$ and (2) for each terminal configuration $\gamma$ such that $\tuple{C,\, G}\to^* \gamma$, $\gamma = \failrm$. In other words, $C$ finitely fails on $G$ if all computations starting from $(C,\, G)$ eventually end in the configuration $\failrm$. 

\begin{figure}[htb]
 \begin{center}
\begin{tabular}{lcl}
$\mathrm{[Call_1]}$ $\frac{\displaystyle G \dder_{\I(R)} H}{\displaystyle\tuple{R,\,G} \to H}$ 
& \hspace{.5em} &
$\mathrm{[Call_2]}$ $\frac{\displaystyle G \not\in \dom(\dder_{\I(R)})}{\displaystyle\tuple{R,\,G} \to \failrm}$
\\\\
$\mathrm{[Seq_1]}$ $\frac{\displaystyle \tuple{P,\, G} \to \tuple{P',\, H}}{\displaystyle \tuple{P;Q,\, G} \to \tuple{P';Q,\, H}}$ 
&&
$\mathrm{[Seq_2]}$ $\frac{\displaystyle \tuple{P,\, G} \to H}{\displaystyle \tuple{P;Q,\, G}\to \tuple{Q,\, H}}$
\\\\
$\mathrm{[Seq_3]}$ $\frac{\displaystyle \tuple{P,\, G} \to \failrm}{\displaystyle \tuple{P;Q,\, G}\to \failrm}$
\\\\
$\mathrm{[If_1]}$ $\frac{\displaystyle \tuple{C,\, G} \to^+ H}{\displaystyle \tuple{\ifte{C}{P}{Q},\, G}\to \tuple{P,\, G}}$
&&
$\mathrm{[If_2]}$ $\frac{\displaystyle \text{$C$ finitely fails on $G$}}{\displaystyle \tuple{\ifte{C}{P}{Q},\, G} \to \tuple{Q,\, G}}$
\\\\
$\mathrm{[Alap_1]}$ $\frac{\displaystyle \tuple{P,\, G} \to^+ H}{\displaystyle \tuple{P!,\, G} \to \tuple{P!,\, H}}$
&&
$\mathrm{[Alap_2]}$ $\frac{\displaystyle \text{$P$ finitely fails on $G$}}{\displaystyle \tuple{P!,\, G} \to G}$
\end{tabular} 
 \end{center}
\caption{Inference rules for core commands}\label{fig:core_sos_rules}
\end{figure}

The concept of finite failure stems from logic programming where it is used to define \emph{negation as failure}\/ \cite{Clark78a}. In the case of GP, we use it to define powerful branching and iteration constructs. In particular, our definition of the if-then-else command allows to ``hide'' destructive tests. 

\begin{example}[Recognizing series-parallel graphs]
\label{ex:series-parallel}
A graph is \emph{series-parallel}\/ if it reduces to a graph consisting of two nodes and an edge between them by the following two operations \cite{Bang_Jensen-Gutin00a,Duffin65a}: (1) Replace a pair of parallel edges by an edge from their source to their target. (2) Given a node $v$ with exactly one incoming edge $e_1$ and exactly one outgoing edge $e_2$ such that the source of $e_1$ and the target of $e_2$ are distinct, replace $e_1$, $e_2$ and $v$ by an edge from the source of $e_1$ to the target of $e_2$.

Suppose that we want to check whether a connected, integer-labelled graph $G$ is series-parallel and, depending on the result, execute either a program $P$\/ or a program $Q$ on $G$. We can do this with the program
\[ \mathtt{main}\; =\; \mathtt{if}\ \{\mathtt{par},\, \mathtt{seq}\}!;\, 
   \mathtt{base}\ \mathtt{then}\ P\ \mathtt{else}\ Q \]
whose rule schemata $\mathtt{par}$, $\mathtt{seq}$ and $\mathtt{base}$ are shown in Figure \ref{fig:series-parallel}. The subprogram $\{\mathtt{par},\, \mathtt{seq}\}!$ applies as long as possible the operations (1) and (2) to the input graph $G$, then the rule schema $\mathtt{base}$ checks if the resulting graph consists of two nodes connected by an edge. Graph $G$ is series-parallel if and only if $\mathtt{base}$ is applicable to the reduced graph. (Note that $\{\mathtt{par},\, \mathtt{seq}\}!$ preserves connectedness and that, by the dangling condition, \texttt{base} is applicable only if the images of its left-hand nodes have degree one.) It is important to note that by the inference rules $\mathrm{[If_1]}$ and $\mathrm{[If_2]}$, the main program executes $P$\/ or $Q$ \emph{on the input graph $G$}\/ whereas the graph resulting from the test is discarded. 
\begin{figure}[htb]
 \begin{center}
  \input{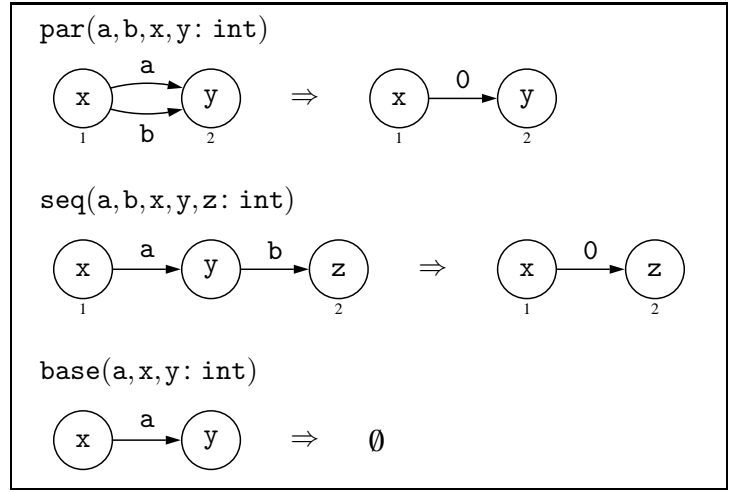}
 \end{center}
\caption{Rule schemata for recognizing series-parallel graphs}
\label{fig:series-parallel}
\end{figure}
\end{example}

The meaning of the remaining GP commands is defined in terms of the meaning of the core commands, see Figure \ref{fig:derived_sos_rules}. We refer to these commands as \emph{derived}\/ commands.

\begin{figure}[hbt]
\begin{center}
\begin{tabular}{lcl}
$\mathrm{[Skip]}$ && $\tuple{\skiptt,\, G} \to \tuple{\mathtt{r},\, G}$\\ && where $\mathtt{r}$ is an identifier for the rule schema $\emptyset \dder \emptyset$
\\[1.5ex]
$\mathrm{[[Fail]}$ && $\tuple{\failtt,\, G} \to \tuple{\{\},\, G}$
\\[1.5ex]
$\mathrm{[If_3]}$ && $\tuple{\ift{C}{P},\, G} \to \tuple{\ifte{C}{P}{\skiptt},\, G}$
\end{tabular} 
\end{center}
\caption{Inference rules for derived commands}\label{fig:derived_sos_rules}
\end{figure}

We can now summarise the meaning of GP programs by a semantic function $\Sem{\_}$ which assigns to each program $P$\/ the function $\Sem{P}$ mapping an input graph $G$ to the set of all possible results of running $P$\/ on $G$. The result set may contain, besides proper results in the form of graphs, the special value $\bot$ which indicates a nonterminating or stuck computation. The \emph{semantic function} $\Sem{\_}\colon \mathrm{ComSeq} \to (\G \to 2^{\G\cup\{\bot\}})$ is defined by\footnote{We write $\Sem{P}G$ for the application of $\Sem{P}$ to a graph $G$.}
   \[\Sem{P}G\, =\, \{H \in \G \mid \tuple{P,\,G}\DSto^+ H\} 
     \cup \{\bot \mid \text{$P$ can diverge or get stuck from $G$}\}\] 
where $P$ \emph{can diverge from} $G$ if there is an infinite sequence $\tuple{P,\,G} \to \tuple{P_1,\,G_1} \to \tuple{P_2,\,G_2} \to \dots$, and $P$ \emph{can get stuck from} $G$ if there is a terminal configuration $\tuple{Q,\,H}$ such that $\tuple{P,\,G} \to^* \tuple{Q,\,H}$.

Note that $\Sem{P}G = \emptyset$ if and only if $P$\/ finitely fails on $G$. In Example \ref{ex:series-parallel}, for instance, we have $\Sem{\{\mathtt{par},\, \mathtt{seq}\}!;\, \mathtt{base}}G = \emptyset$ for every connected graph $G$\/ containing a cycle. This is because the graph resulting from $\{\mathtt{par},\, \mathtt{seq}\}!$ is still connected and cyclic, so the rule schema \texttt{base} is not applicable. 

A program can get stuck only in two situations: either it contains a subprogram $\ifte{C}{P}{Q}$ where $C$ both can diverge from some graph and cannot produce a proper result from that graph, or it contains a subprogram $B!$ where the loop's body $B$ possesses the said property of $C$. The evaluation of these subprograms will get stuck because the inference rules for branching and iteration are not applicable.

\section{Conclusion}
\label{sec:conclusion}

GP is an experimental rule-based language for high-level problem solving in the domain of graphs, freeing programmers from handling low-level data structures. The hallmark of GP is syntactic and semantic simplicity. Conditional rule schemata for graph transformation allow to express  application conditions and computations on labels, in addition to structural changes. The semantics of rule schemata is orthogonal to the semantics of control constructs, making it easy to change the format of rules or graphs.

The operational semantics of programs describes the effect of GP's control constructs in a natural way and captures the nondeterminism of the language. In particular, powerful branching and iteration commands have been defined using the concept of finite failure. Destructive tests on the current graph can be hidden in the condition of the branching command, and nested loops can be coded since arbitrary subprograms can be iterated as long as possible.

Future extensions of GP may include recursive procedures for writing complex algorithms (see \cite{Steinert07a}), and a type concept for restricting the shape of graphs. Our goal is to support formal reasoning on graph programs by developing static analyses for properties such as termination and confluence (uniqueness of results), and a calculus and tool support for program verification.

\bibliographystyle{plain}
\bibliography{/usr/det/Bibtex/abbr,%
              /usr/det/Bibtex/absredu.bib,%
              /usr/det/Bibtex/trs,%
              /usr/det/Bibtex/gragra,%
              /usr/det/Bibtex/graph-algorithms,%
              /usr/det/Bibtex/graphs-misc,%
              /usr/det/Bibtex/gratra-languages,%
              /usr/det/Bibtex/proglang,%
              /usr/det/Bibtex/logicprog,%
              /usr/det/Bibtex/misc}

\end{document}